# From Awareness to Application: Strengthening Recruitment for NSF S-STEM Scholarships in Computer Science

Xiaohui Yuan
Department of Computer Science and Engineering
University of North Texas
Denton, Texas, USA

# From Awareness to Application: Strengthening Recruitment for NSF S-STEM Scholarships in Computer Science

**Abstract**
Recruiting academically strong but financially disadvantaged students into NSF S-STEM scholarship programs remains a persistent challenge in computer science (CS) education, particularly among underrepresented populations. This paper presents the design and initial implementation of a suite of targeted recruitment strategies for our NSF S-STEM funded competitive scholarship project, which supports undergraduate CS students who are academically strong yet financially disadvantaged. Our recruitment strategy leverages multiple channels, including departmental communications, faculty referrals, targeted digital campaigns, presentations in gateway courses, and collaborations with the scholarship department that supports students facing financial challenges. Information sessions and early outreach efforts were employed to increase awareness and reduce perceived barriers to applying. Data from our recruitment includes applicant demographics, academic performance, financial aid profiles, recruitment source tracking, and survey responses on students' awareness and decision-making processes. This data provides a foundation for evaluating the reach and effectiveness of various recruitment strategies and identifying factors that influence student application decisions. Quantitative and qualitative research approaches are employed to examine the implementation and outcomes of proactive recruitment strategies. Our preliminary analysis indicates that direct information sessions and departmental emails are effective recruitment strategies, accounting for a large portion of eligible applications. Our findings emphasize the importance of early communication about the program, clearly defined eligibility criteria, and a streamlined application process. Future work will involve analysis of recruitment channel effectiveness, longitudinal tracking of applicant outcomes, and refinement of messaging strategies. By sharing ongoing progress and lessons learned from our project, this paper contributes evidence-based insights into recruitment practices and offers strategies that can be adapted by other institutions implementing NSF S-STEM programs.

## 1. Introduction

Artificial Intelligence (AI) has become a foundational component of modern Computer Science (CS) education, reshaping curricular priorities and redefining workforce expectations. National and federal workforce strategies increasingly emphasize the need to prepare an AI-ready workforce by expanding access to high-quality CS education and ensuring that academically talented students are not excluded due to financial constraints [10, 11, 12]. However, despite growing demand for AI expertise, financial barriers continue to limit participation and persistence in CS programs, particularly among students from underrepresented, low-income, and first-generation backgrounds.

The Department of Computer Science and Engineering (CSE) at the University of North Texas (UNT) serves a highly diverse student population, many of whom demonstrate strong academic potential but experience significant financial need. In recent years, requests for financial aid and emergency assistance have increased sharply, reflecting broader national trends in housing insecurity, food insecurity, and economic instability among college students [13]. These challenges disproportionately affect minority and first-generation students and have been shown to negatively impact academic performance, persistence, and overall well-being [14]. The COVID-19 pandemic further exacerbated these vulnerabilities, intensifying economic hardship and widening inequities in access to educational opportunities [15]. As a result, targeted financial and academic support mechanisms have become increasingly critical to ensuring student success in CS.

NSF S-STEM programs are designed to address these challenges by supporting academically strong, financially disadvantaged students through a combination of financial assistance, mentoring, and academic support. Our team received an NSF S-STEM award to support a cohort of undergraduate CS students, referred to as *AI Scholars*, to reduce financial burden so that students can focus on coursework, research engagement, and professional development rather than excessive employment. The program is intended to improve retention, academic success, and graduation rates among high-achieving, low-income CS majors.

Despite the availability of substantial funding and institutional commitment, recruiting eligible students into S-STEM programs remains a persistent and underexplored challenge. Many qualified students are unaware of scholarship opportunities, misunderstand eligibility requirements, or perceive the application process as inaccessible or intimidating. These barriers are particularly pronounced for underrepresented and first-generation students, who may lack familiarity with federally funded programs or distrust institutional initiatives. Additionally, recruitment efforts are often hindered by late outreach, fragmented communication across departments, and limited coordination among faculty, advisors, financial aid offices, and scholarship administrators. As a result, S-STEM programs frequently struggle to reach the very populations they are designed to serve.

This paper addresses this critical gap by presenting the design and initial implementation of a comprehensive, multi-channel recruitment strategy tailored to academically talented, financially disadvantaged CS students. Our approach integrates departmental communications, faculty referrals, targeted digital outreach, presentations in gateway CS courses, information sessions, and close collaboration with campus scholarship and financial aid offices. By emphasizing early outreach, clear eligibility messaging, trust-building mechanisms, and cross-unit coordination, our recruitment model seeks to increase awareness, reduce perceived barriers, and encourage qualified students to apply. The strategies described in this paper offer a scalable and replicable framework for improving recruitment effectiveness in NSF S-STEM programs and advancing equity in CS education.

The rest of this article is organized as follows: Section 2 reviews the related work of student recruitment, especially related to the NSF S-STEM programs. Section 3 presents our method. Section 4 discusses the results. Section 5 concludes this paper with a summary.

## 2. Related Work

Broadening participation for academically strong yet financially disadvantaged students has long been a central priority in STEM education policy and research, with the NSF Scholarships in Science, Technology, Engineering, and Mathematics (S-STEM) program serving as one of the primary federal mechanisms for advancing this goal. S-STEM initiatives are intentionally designed to reduce financial barriers while pairing scholarships with evidence-based academic and co-curricular supports, including mentoring, advising, cohort-building, and undergraduate research experiences. Recent national evaluations of S-STEM programs consistently demonstrate positive impacts on student persistence, academic performance, and graduation outcomes; however, they also emphasize that financial aid alone is insufficient to achieve these outcomes without complementary structures that foster belonging and academic integration [1, 2]. Multi-institutional S-STEM studies report gains in retention and sense of community among scholars, yet repeatedly identify recruitment as a persistent bottleneck, particularly because many eligible students do not recognize themselves as competitive applicants or do not encounter program information early enough to act on it [1].

Systematic reviews of diversity-focused STEM interventions further reinforce these findings. Palid et al. [3] synthesize over a decade of research on programs supporting underrepresented and low-income STEM students and identify early outreach, mentoring, and proactive advising as consistent elements of successful interventions. At the same time, the review highlights a notable limitation in the literature: recruitment strategies are frequently described at a high level or anecdotally, with little empirical evaluation of how students are reached or which channels are most effective in motivating applications. This lack of recruitment-focused evidence limits the transferability and scalability of otherwise successful scholarship models.

National S-STEM reports and program guidance further underscore the importance of early communication, coordinated institutional messaging, and structured student support in strengthening scholar engagement and persistence [1]. Within this broader context, the UNT AI Scholars S-STEM program aligns closely with NSF priorities by integrating renewable financial support with mentoring, cohort-based activities, and research engagement for low-income CS students. While formal evaluation of the program's recruitment strategies is ongoing, the UNT initiative provides an illustrative case of how national S-STEM goals are operationalized within a large, diverse, high-need public research institution. Importantly, the program's experience mirrors national findings that recruitment remains one of the most challenging aspects of S-STEM implementation.

Research across STEM and engineering education increasingly suggests that multi-channel recruitment approaches are essential for reaching the underserved student population. Studies have documented the effectiveness of combining targeted digital outreach, classroom-based presentations, faculty referrals, partnerships with advising and financial aid offices, and collaborations with student support units in identifying and encouraging eligible students to apply [4, 5]. Faculty members, in particular, are consistently identified as trusted messengers whose encouragement plays a critical role in students' decisions to pursue competitive scholarships or enrichment opportunities. At the same time, prior research demonstrates that low-income and first-generation students are often discouraged by unclear eligibility criteria, complex application requirements, and uncertainty about whether they "belong" in selective programs [6]. Recommended practices include simplified eligibility messaging, early outreach in gateway courses, information sessions, and personalized advising support. Despite these recommendations, few studies offer data-driven comparisons of recruitment channels or systematically analyze how different outreach mechanisms influence application behavior, especially within S-STEM-funded CS programs. Institutional reports from peer S-STEM projects also reference strategies such as high school engagement and community college partnerships to broaden applicant pools, but these approaches are rarely evaluated using empirical recruitment data.

Computing education research provides additional insight into why recruitment challenges are particularly pronounced in CS. Studies of the computing "leaky pipeline" show that barriers to participation emerge early and are shaped by institutional messaging, access to social capital, and faculty engagement, disproportionately affecting students from underrepresented and low-income backgrounds [7]. Relationship-based and cohort-oriented interventions have been shown to improve students' sense of belonging, self-efficacy, and persistence in computing programs [8], suggesting that recruitment should be viewed not merely as an entry point but as a foundational component of long-term student success. Research on culturally responsive mentoring and inclusive program design further emphasizes that early contact with faculty and scholarship programs can help students envision themselves as capable and valued participants in computing pathways [9]. These findings reinforce NSF S-STEM priorities that frame recruitment as an equity-driven process embedded within a broader ecosystem of academic, financial, and social support.

Collectively, the literature demonstrates that while S-STEM programs are well documented in terms of student outcomes and support structures, there remains a significant gap in empirical research focused specifically on recruitment effectiveness, particularly within CS programs serving financially disadvantaged and underrepresented students. Existing studies seldom disaggregate recruitment channels, track how applicants learn about opportunities, or analyze how perceptions of eligibility and accessibility influence application behavior.

This study addresses this gap by providing a data-driven analysis of targeted recruitment strategies within an NSF S-STEM-funded CS scholarship program at a large public research university. By examining applicant demographics, academic and financial profiles, recruitment source tracking, and survey responses, this work offers empirical evidence on the relative effectiveness of faculty referrals, departmental communications, and early outreach efforts. The findings contribute actionable guidance for institutions seeking to strengthen recruitment pipelines for high-achieving, financially disadvantaged CS students and advance broader engineering education goals centered on equity, access, and inclusive participation.

## 3. Method

This study employed a structured, multi-channel recruitment methodology designed to directly address persistent challenges in recruiting academically strong yet financially disadvantaged students into NSF S-STEM programs. The minimum criteria used for selecting students for the scholarships are in Table 1. Our primary focus is on sophomore students, who, at the time of application, are in their first year at UNT. We are also open to students who are going to junior or even senior level.

Table 1. Criteria for the scholarship

- U.S. Citizens or nationals of the U.S.
- Enrolled at least half-time in CSE programs
- Low-income and demonstrated unmet financial need
- Minimum a GPA of 3.0

The recruitment strategy was implemented during the Spring 2025 semester as part of the initial recruitment cycle for the UNT AI Scholars S-STEM program. The UNT AI Scholars program is housed in the Department of CSE at UNT, a large public research university serving a diverse and high-need student population. Recruitment efforts focused on undergraduate CS majors who demonstrated strong academic performance and unmet financial need, with particular attention to first-generation, transfer, and historically underrepresented students. The recruitment design intentionally emphasized early academic pathways where intervention has the greatest potential to influence persistence and degree completion.

Recruitment activities began in Spring 2025 with an intentional shift away from passive, email-only notifications toward embedded, relationship-based outreach. Scholarship information was integrated directly into high-enrollment "gateway" CS courses, including CSCE 1030, CSCE 1040, and CSCE 1015, which are recognized institutional inflection points for retention and progression in the CS major.

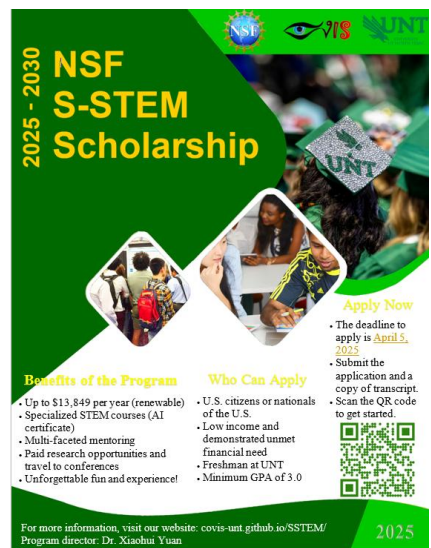

Figure 1. NSF S-STEM at UNT program flyer.

During the in-class presentations (~10 minutes), members of the project team introduced the S-STEM program, clarified eligibility criteria using student-facing language, and emphasized the program's academic and professional benefits. To reduce procedural friction, students were provided immediate access to application resources through QR codes and short links embedded in presentation slides, allowing them to explore eligibility and application steps in real time. This approach ensured that information reached students in familiar academic settings and lowered barriers to initial engagement.

To address documented trust gaps among first-generation and historically underrepresented students, faculty were positioned as a primary program messenger. Faculty teaching gateway and mid-level courses reinforced scholarship messaging during lectures and advising interactions, providing social validation and legitimacy to the opportunity. Rather than framing the S-STEM program solely as financial aid, faculty emphasized its competitive and cohort-based nature, presenting it as an elite academic fellowship designed to recognize and invest in student potential. This reframing helped counter students' tendencies to underestimate their competitiveness and fostered a sense of belonging within the CS academic community. Faculty endorsements served as a critical catalyst for application, particularly for students who may not have otherwise self-identified as scholarship candidates. Personalized outreach from trusted faculty or advisors, including messages explicitly stating that a student was a strong candidate, was used to boost self-efficacy and encourage application completion. These high-trust, data-informed referrals substantially increased both the volume and quality of applicants.

Classroom engagement was reinforced through a coordinated, multi-channel communication strategy designed to provide repeated, consistent touchpoints. Departmental email announcements were synchronized with posts on the Canvas learning management system, while digital signage in computing laboratories and student commons provided ongoing visual reminders. Targeted social media outreach extended recruitment visibility into students' preferred online spaces. Recurring virtual and in-person information sessions were held from mid-February through March, offering multiple opportunities for students to engage with program staff, ask questions, and build familiarity with the scholarship. These sessions emphasized transparency in eligibility, application expectations, and program benefits, further strengthening trust and understanding.

To prevent fragmented communication and missed recruitment opportunities, the project team established a formal collaboration network among the CS department, academic advising, and the Office of Financial Aid. A recruitment calendar ensured consistency across outreach channels, while regular interdepartmental briefings aligned timelines and responsibilities. Internal referral protocols enabled financial aid officers to identify and flag high-performing students with unmet financial need, feeding directly into the S-STEM recruitment pipeline. Faculty and professional advisors served as frontline identifiers of student potential and were provided with streamlined nomination tools and simplified eligibility indices. These tools emphasized not only GPA and financial criteria but also indicators of resilience, curiosity, and collaborative aptitude observed in classroom and laboratory settings.

Our recruitment activities were concentrated in the early weeks of the semester, before the peak of midterm workloads and before students finalized their employment commitments. Outreach efforts were also aligned with FAFSA submission periods, transfer student orientations, and key advising milestones to reach students during moments of heightened financial and academic decision-making. Special emphasis was placed on sophomore-level "bottleneck" courses, where students had already demonstrated academic perseverance and were positioned to benefit most from longitudinal S-STEM support. This timing strategy addressed documented challenges related to late outreach and reduced student capacity to engage with scholarship opportunities.

Recognizing that procedural barriers often deter eligible students, the project team implemented structured application support mechanisms. Collaborative application workshops were hosted to provide step-by-step guidance on crafting personal statements, requesting letters of recommendation, and navigating submission requirements. Workshops emphasized narrative development that highlighted academic merit, resilience, and motivation, particularly for students with limited prior exposure to competitive scholarship environments.

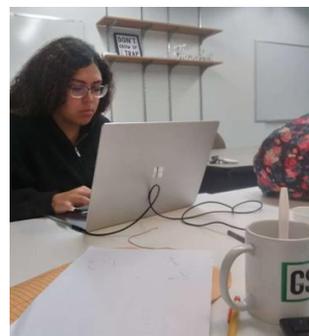

Students received coaching on professional communication with faculty recommenders and were offered iterative feedback on application drafts from peers and program coordinators, as shown in Figure 2. By transforming the application process into a guided professional development experience, the program reduced inequities in access to social capital and ensured that submitted materials accurately reflected students' potential.

Figure 2. Application consultation.

All recruitment activities were designed with evaluation in mind. Recruitment source tracking, applicant demographic and academic data, and survey responses were collected to enable systematic analysis of recruitment effectiveness. This design allows for disaggregation of recruitment channels and provides empirical evidence addressing a key gap in the S-STEM literature: understanding how students learn about scholarship opportunities and what motivates them to apply.

## 4. Results and Discussion

In the Spring 2025 recruitment cycle of the UNT NSF S-STEM program, we received over 130 inquiries from students, among which a total of 107 students submitted applications. Table 2 presents the statistics of the application pool. Within CS, the largest proportion of applicants were early-stage students (1-2 semesters completed), totaling 30 applicants, followed by 19 students at the 3-4 semester level and 15 students with five or more semesters completed. A similar pattern is observed across CE and Cybersecurity, where early-stage applicants outnumber those at more advanced levels. In contrast, IT applicants were more evenly distributed across academic levels, with a slight concentration among students with five or more semesters completed.

Table 2. Summary of the applicants (2025)

| Program | CS | | CE | | Cyber | | IT | |
|---|---|---|---|---|---|---|---|---|
| Level (semesters) | 1-2 | 30 | 1-2 | 10 | 1-2 | 7 | 1-2 | 3 |
| | 3-4 | 19 | 3-4 | 6 | 3-4 | 4 | 3-4 | 2 |
| | 5+ | 15 | 5+ | 4 | 5+ | 3 | 5+ | 4 |

The composition of the application pool provides important evidence regarding the effectiveness and equity of the recruitment strategy. The strong representation of early-stage students across CS, CE, and Cybersecurity suggests that embedding outreach within gateway courses and aligning recruitment with early academic milestones successfully reached students at critical inflection points for retention and persistence. This outcome directly addresses a well-documented challenge in S-STEM recruitment, where outreach often occurs too late to meaningfully influence students' academic trajectories. The relatively smaller number of advanced-level applicants (5+ semesters) may reflect competing demands such as employment, capstone requirements, or impending graduation, which are common barriers for late-stage students. This pattern reinforces the importance of early, proactive recruitment and longitudinal support, particularly for financially disadvantaged students who benefit most from sustained scholarship funding.

Table 3 lists the qualified applicants after the initial screening. Among the 46 qualified applicants, the majority (69.6%) were CS majors, which is consistent with the enrollment of our four undergraduate programs. However, a substantial portion of the applicant pool came from closely related computing disciplines, including Computer Engineering (13.0%), Cybersecurity (10.9%), and Information Technology (6.5%). Qualified students are fairly evenly distributed across academic levels in CS, while other programs show more concentration at early stages. This distribution indicates that recruitment efforts extended beyond a single major and reached a broader computing student population.

The average GPA indicates that qualified students are academically strong, with averages consistently above 3.4. This confirms that recruitment strategies are successfully targeting the intended population of academically talented yet financially disadvantaged students. Many early-stage students fall within the 10-20k income bracket, while mid- and upper-level students are more evenly distributed across higher brackets. The pool composition demonstrates that targeted, multi-channel recruitment effectively attracts academically strong students who meet both merit and financial criteria. Representation across academic levels and disciplines suggests that the recruitment strategy supports inclusion while ensuring program rigor. The presence of students in high-need brackets across all levels underscores the potential impact of scholarships in improving retention, persistence, and long-term academic success.

Table 3. Summary of qualified applicants (2025)

| Program | Level (semesters) | | GPA | Income | |
|---|---|---|---|---|---|
| CS | 1-2 | 11 | 3.4 | 10-20 | 10 |
| | 3-4 | 10 | | 20-40 | 10 |
| | 5+ | 11 | | 40+ | 12 |
| CE | 1-2 | 2 | 3.8 | 10-20 | 4 |
| | 3-4 | 3 | | 20-40 | 1 |
| | 5+ | 1 | | 40+ | 1 |
| Cyber | 1-2 | 2 | 3.4 | 10-20 | 1 |
| | 3-4 | 3 | | 20-40 | 1 |
| | 5+ | 0 | | 40+ | 3 |
| IT | 1-2 | 1 | 3.5 | 10-20 | 1 |
| | 3-4 | 1 | | 20-40 | 0 |
| | 5+ | 1 | | 40+ | 2 |

The statistics of the qualified pool highlight the effectiveness of strategies aimed at reducing eligibility misperceptions and procedural barriers. By embedding scholarship outreach in courses, leveraging trusted faculty as messengers, and providing structured application support, the program successfully identified students with both high academic potential and significant financial need. The data suggests that continued emphasis on early outreach and multi-department coordination can further strengthen the pipeline for underrepresented and low-income computing students.

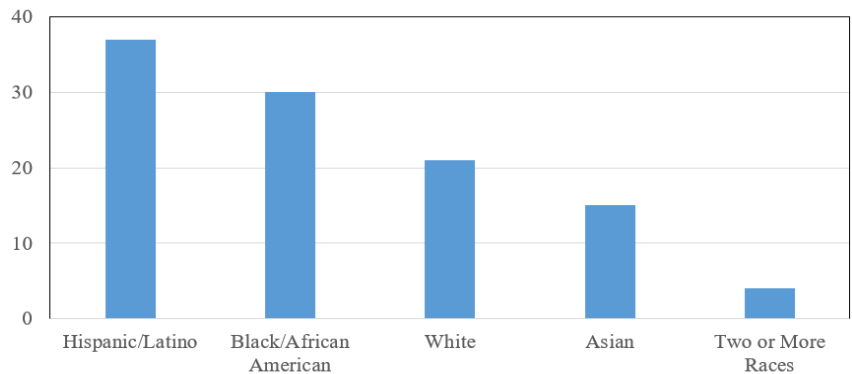

Figure 3. Distribution of applicants by ethnicity

Figure 3 depicts the demographic composition of the applicant pool, which reflects strong participation from historically underrepresented groups in computing. As shown in Table 3, Hispanic/Latino (34.6%) and Black/African American (28.0%) students together comprised over 60% of applicants. White (18.7%) and Asian (14.0%) students made up smaller portions of the pool, with the remaining applicants identifying as multiracial or other ethnicities. This demographic distribution aligns closely with the equity goals of the NSF S-STEM program and suggests that the targeted recruitment strategies were effective in reaching students from underrepresented backgrounds.

The distribution of applicants by recruitment venue is illustrated in Figure 4. Faculty referrals accounted for the largest share of applicants (28.0%), followed by gateway course presentations (21.5%). Together, these two high-touch, relationship-based recruitment channels generated nearly half of all applications. Lower-touch channels, such as departmental emails (11.2%), information sessions (11.2%), Canvas announcements (9.3%), and financial aid or scholarship office referrals (8.4%), contributed additional applicants, while peer referrals represented a smaller share (3.7%). These results underscore the importance of embedding recruitment within academic contexts and leveraging trusted faculty messengers. While digital and administrative channels played a supportive role, personalized outreach and in-class engagement were the most effective mechanisms for motivating eligible students to apply. Students reported different levels of understanding by recruitment venue. Although the departmental email accounted for the largest portion of the responses, many applicants confirmed that they received the program information via different channels, and information sessions and presentations in the gateway courses motivated their applications. This highlights the critical role of faculty endorsement in reducing psychological barriers and reinforcing students' perceptions of belonging and competitiveness. Faculty-mediated recruitment appears particularly effective in addressing the misperceptions of eligibility that disproportionately affect first-generation and low-income students.

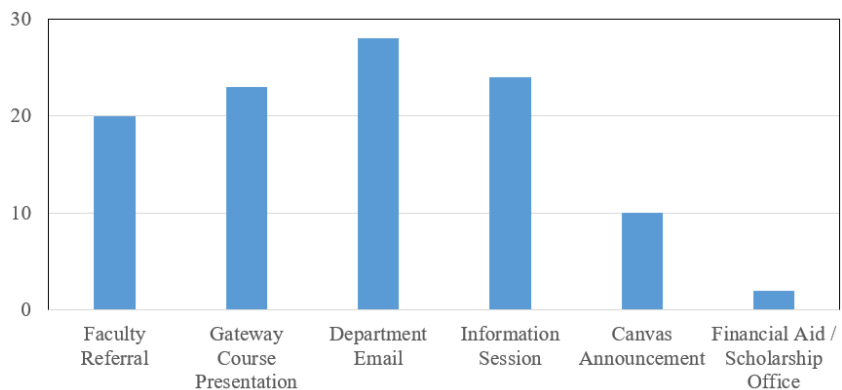

Figure 4. Distribution of applicants by recruitment channel.

In addition, the project team offered individual application assistance and workshops focused on personal statement development and submission logistics. The volume of applications indicates that guided application support can function as an equity-oriented intervention, enabling students with limited prior exposure to competitive scholarships to submit strong applications that accurately reflect their academic potential.

## 5. Conclusion

Recruiting academically strong yet financially needy students into NSF S-STEM scholarship programs remains a persistent challenge. This study presents a multi-channel recruitment strategy implemented for the UNT S-STEM program, designed to address key barriers, including low awareness, eligibility misperceptions, fragmented communication, and timing constraints.

Our results suggest that embedding scholarship outreach into high-enrollment gateway courses, leveraging faculty as messengers, coordinating multi-departmental communications, and providing structured application support can effectively engage eligible students. Early-stage students, particularly in CS, were successfully recruited, while meaningful representation from CE, Cybersecurity, and IT programs suggests that interdisciplinary outreach strategies broaden program impact. The qualified applicant pool reflects both high academic achievement (GPA $\geq$ 3.4) and verified financial need, demonstrating that our recruitment efforts reached the population intended to benefit most from S-STEM support. Analysis of recruitment channels revealed that departmental emails, information sessions, faculty referrals, and gateway course presentations were the most effective mechanisms. Structured workshops and individual application consulting mitigated procedural barriers, allowing students with limited prior experience in competitive scholarship applications to submit strong, reflective materials. These strategies contributed to a diverse and academically talented applicant pool, supporting the program's goals of improving retention, persistence, and long-term success.

This study contributes empirical evidence to a gap in the literature on recruitment effectiveness within NSF S-STEM programs, particularly in CS and related disciplines. The results provide actionable guidance for institutions seeking to expand access to financial aid and mentorship programs for low-income, underrepresented students, emphasizing early engagement, faculty advocacy, coordinated institutional efforts, and equity-driven design. Future work will examine longitudinal outcomes for scholars, evaluate the relative impact of specific recruitment channels on retention and graduation, and explore the scalability of the framework across institutions and STEM disciplines. In summary, a deliberate, data-informed, and relationship-based recruitment approach not only increases the quantity and quality of applicants but also fosters equity and inclusion, helping ensure that financially disadvantaged students with strong academic potential have the resources and support needed to succeed in computing pathways.